\documentclass[lettersize,journal]{IEEEtran}
\usepackage{amsmath,amsfonts}
\usepackage{mathtools}
\usepackage{array}
\usepackage[caption=false,font=normalsize,labelfont=sf,textfont=sf]{subfig}
\usepackage{textcomp}
\usepackage{stfloats}
\usepackage{url}
\usepackage{verbatim}
\usepackage{graphicx}
\usepackage{subcaption}
\usepackage{cite}
\usepackage{pifont}

\newcommand{\cmark}{\ding{51}}%
\usepackage{multirow}
\usepackage{multicol}

\usepackage{mathrsfs}
\usepackage{algorithm}
\usepackage{algpseudocode}
\makeatletter
\algrenewcommand\ALG@beginalgorithmic{\footnotesize}
\makeatother

\usepackage{braket}

\usepackage{acro}

\usepackage{listings}
\usepackage{xcolor}
\usepackage{soul} 
\definecolor{eclipseStrings}{RGB}{0.0,0.0,0.0}
\definecolor{eclipseKeywords}{RGB}{0.0,0.0,0.0}
\colorlet{numb}{magenta!60!black}
\lstdefinelanguage{json}{
    basicstyle=\normalfont\ttfamily\small,
    commentstyle=\color{eclipseStrings}, 
    stringstyle=\color{eclipseKeywords}\footnotesize, 
    numberstyle=\scriptsize,
    stepnumber=1,
    numbersep=8pt,
    showstringspaces=false,
    breaklines=true,
    frame=lines,
    backgroundcolor=\color{white}, 
    string=[s]{"}{"},
    comment=[l]{:\ "},
    morecomment=[l]{:"},
    literate=
        *{0}{{{\color{numb}0}}}{1}
         {1}{{{\color{numb}1}}}{1}
         {2}{{{\color{numb}2}}}{1}
         {3}{{{\color{numb}3}}}{1}
         {4}{{{\color{numb}4}}}{1}
         {5}{{{\color{numb}5}}}{1}
         {6}{{{\color{numb}6}}}{1}
         {7}{{{\color{numb}7}}}{1}
         {8}{{{\color{numb}8}}}{1}
         {9}{{{\color{numb}9}}}{1}
}

\begin{document}

\title{RSLAQ - A Robust SLA-driven 6G O-RAN QoS xApp using deep reinforcement learning}

\author{Noe M. Yungaicela-Naula, Vishal Sharma, Sandra Scott-Hayward
\thanks{The authors are with the Centre for Secure Information Technologies (CSIT),
Queen's University Belfast,
Belfast, Northern Ireland, UK.\\ E-mail: n.yungaicela@qub.ac.uk}
\thanks{This work has been submitted to IEEE for possible publication April, 2025.}}

\markboth{Journal of \LaTeX\ Class Files,~Vol.~14, No.~8, August~2021}%
{Shell \MakeLowercase{\textit{et al.}}: ML for Reliable QoS framework for 6G O-RAN}


\maketitle

\begin{abstract}
The evolution of 6G envisions a wide range of applications and services characterized by highly differentiated and stringent Quality of Service (QoS) requirements. Open Radio Access Network (O-RAN) technology has emerged as a transformative approach that enables intelligent software-defined management of the RAN. A cornerstone of O-RAN is the RAN Intelligent Controller (RIC), which facilitates the deployment of intelligent applications (xApps and rApps) near the radio unit. In this context, QoS management through O-RAN has been explored using network slice and machine learning (ML) techniques. Although prior studies have demonstrated the ability to optimize RAN resource allocation and prioritize slices effectively, they have not considered the critical integration of Service Level Agreements (SLAs) into the ML learning process. This omission can lead to suboptimal resource utilization and, in many cases, service outages when target Key Performance Indicators (KPIs) are not met. This work introduces RSLAQ, an innovative xApp designed to ensure robust QoS management for RAN slicing while incorporating SLAs directly into its operational framework. RSLAQ translates operator policies into actionable configurations, guiding resource distribution and scheduling for RAN slices. Using deep reinforcement learning (DRL), RSLAQ dynamically monitors RAN performance metrics and computes optimal actions, embedding SLA constraints to mitigate conflicts and prevent outages. Extensive system-level simulations validate the efficacy of the proposed solution, demonstrating its ability to optimize resource allocation, improve SLA adherence, and maintain operational reliability  ($>95\%$) in challenging scenarios.

\end{abstract}


\section{Introduction}
\IEEEPARstart{5}{G/6G} introduces services with stringent Quality of Service (QoS) requirements, such as enhanced mobile broadband (eMBB), ultra-reliable low-latency communications (URLLC), and machine-type communications (MTC). Operating these services is highly challenging, as real deployments can contain multiple instances of these services {with diverse QoS requirements}. In the radio access network (RAN), this challenge scales with dynamic user behavior, limited radio resources, and fast channel condition variations.  

{RAN slicing has been considered one of the most promising techniques to manage the QoS in 5G/6G} \cite{Kibeom2023TechnologyTrend}. {In a sliced RAN, Service Level Agreements (SLAs) are used between the network operator and the tenants to declare expected performance and quality of the services details} \cite{ORANWG12023UCAReport}. {Requirements of the service instances are specified in terms of key performance indicators (KPIs), such as throughput and latency. Violation of SLAs can have severe consequences for the network operator, such as financial penalties,  reputational damage, and more critically, legal and regulatory implications (e.g., in sectors involving critical services like healthcare).}

{To maximize their revenues and avoid SLA violations, network operators must allocate RAN resources efficiently, dynamically, and in an automated manner.  Open RAN (O-RAN) offers a solution enabling dynamic resource management through the RAN Intelligent Controller (RIC).  O-RAN helps translate the high-level requirements expressed by the operator in SLAs into low-level network parameters (KPIs). Furthermore, an intelligent application (xApp) within the RIC can help keep the KPIs of different slices near the target KPIs (SLAs). The xApp can utilize machine learning (ML) and real-time metrics from the RAN and user behavior to optimize resource allocation per slice. The xApp must guarantee that the slices operate independently, i.e., the performance in one slice does not negatively influence the
performance of other slices. Furthermore, the slices can have competing or conflicting QoS requirements. Lastly, robustness is required in SLA management to guarantee that the SLAs are reliably maintained in different scenarios, e.g., varying traffic.} 
 
{In practice, SLA management with O-RAN presents a real challenge} \cite{sghaier2023review}. {Therefore,} the latest approaches to RAN slicing have leveraged deep reinforcement learning (DRL) to optimize resources while meeting QoS requirements. However, these approaches focuse {on resource} optimization, such as providing high throughput for eMBB users and low latency for URLLC users. While some of these approaches address QoS conflicts through prioritization, they often fail to fully model the broader network operator intents expressed in SLAs, which require finer granularity.  This limitation results in two primary issues: (1) suboptimal resource utilization; and (2) reduced reliability in QoS compliance. For example, if the throughput for an eMBB slice is being maximized and exceeds the agreed SLA, the surplus resources could be reallocated to other slices that are falling short of their SLAs or released to enhance energy efficiency in the system. Moreover, if reliability  is not considered, using only priorities to decide the distribution of resources for QoS operations can lead to resource starvation in slices with lower priority, potentially causing outages ({KPIs fall below the agreed thresholds}). In 6G, where most {slice} types demand high reliability, such {outages are unacceptable}.

By modeling the distinct SLA requirements of the slices in a more granular manner and incorporating them into the DRL learning process, a better balance of resource allocation can be achieved, optimizing resources while also meeting the target key performance indicators (KPIs). In this work, we propose RSLAQ, an xApp that reliably controls the QoS of different services in the RAN. The xApp receives policies from  {the operator through the RIC} as target KPIs and uses a DRL agent to guarantee that these targets are met while optimizing the resource share between different services. First, a general model is presented, followed by an evaluation of a specific design for operating three slices: eMBB, URLLC, and MTC.

The contributions of this work are as follows:

\begin{itemize}
    \item We provide insights into the existing strategies for operating the QoS in 3GPP 5G NG-RAN and O-RAN through a comprehensive review of recent studies from academia and industry.
    
    \item We {present} RSLAQ, a DRL-based xApp for the reliable operation of QoS RAN that {adheres to 3GPP and O-RAN specifications}. This design incorporates granular SLA requirements as target KPIs, integrating them into the learning process of the DRL agent.

    \item  We demonstrate the effectiveness of our solution by using highly-detailed system-level simulations {that follow 3GPP specifications}.
\end{itemize}

The rest of this document is organized as follows: Section~\ref{sec:2} presents the background on QoS operations within 3GPP and O-RAN and the existing ML-based strategies. The architecture of the system proposed in this study is detailed in Section~\ref{sec:3}. Section \ref{sec:4} presents RSLAQ, our QoS xApp that includes the design of the DRL approach. Experimental results are reported in Section~\ref{sec:5}. Discussion and conclusions are provided in Sections~\ref{sec:6} and \ref{sec:7}, respectively.

\section{Background and State of the art}\label{sec:2}

This section presents a comprehensive review of  QoS operation in 3GPP 5G and O-RAN.

\subsection{3GPP QoS framework}

In the 3GPP 5G framework, the Access and Mobility Management Function (AMF) and the User Plane Function (UPF) (which are in the 5GC) filter the traffic from different services and map them to QoS flows (using QoS identifiers, QFIs) based on agreed service requirements. Within the RAN, the Service Data Adaptation Protocol (SDAP) associates QFIs with {data radio bearers (DRBs)} managed by the RLC/MAC protocols \cite{3GPPTS23501}. The MAC layer then schedules resources using queue management algorithms such as Round Robin (\textit{RR}), Proportional Fairness (\textit{PF}), or Best Quality Channel Indicator (\textit{BCQI})~\cite{Cox2020Introduction}.

The QoS strategy used in the 3GPP 5G QoS framework is semi-dynamic because the DRBs are set up in a pre-determined and fixed manner. Furthermore, previous studies have highlighted a limitation regarding flow granularity \cite{Irazabal2024TCRAN}. Despite the 5GC supporting relatively high granularity with up to 64 QFIs per UE, the 5G NR restricts the number of possible DRBs per UE to 32. This restriction can complicate the prioritization of similar flows, potentially causing data starvation for some flows. Finally, in the 3GPP 5G QoS framework, application requirements and channel variations are not closely integrated with RAN resource optimization mechanisms. Typical MAC schedulers utilize physical and logical channel metrics, but are generally unaware of QoS requirements. For example, \textit{PF} aims to maximize cell throughput by considering channel quality indicators (CQI) while ensuring a minimum level of service for all users. \textit{RR} scheduler distributes resources equally among users. \textit{BCQI} prioritizes users with the highest CQI for transmission, {maximizing cell throughput}. Variations of these algorithms, such as weighted proportional fairness, can prioritize services during scheduling. However, they are only capable of reacting to short-term changes, rely on local information, and do not account for all intentions of the network operator. These limitations of 3GPP are being addressed by O-RAN technology, as detailed next.

\subsection{O-RAN assisted QoS} \label{sec:oranassistedQoS}
{In O-RAN, the RAN is split into the central unit (CU), the distributed unit (DU), and the radio unit (RU).  These components expose E2 service models (E2SMs) to the RIC.  The RIC uses both the Non-RT RIC and the Near-RT RIC to manage RAN resources with millisecond and second-level granularity, respectively} \cite{ORANWG12023Architecture, Sirotkin20205G}.

\begin{figure}[!t]
\centering
\includegraphics[width=\linewidth]{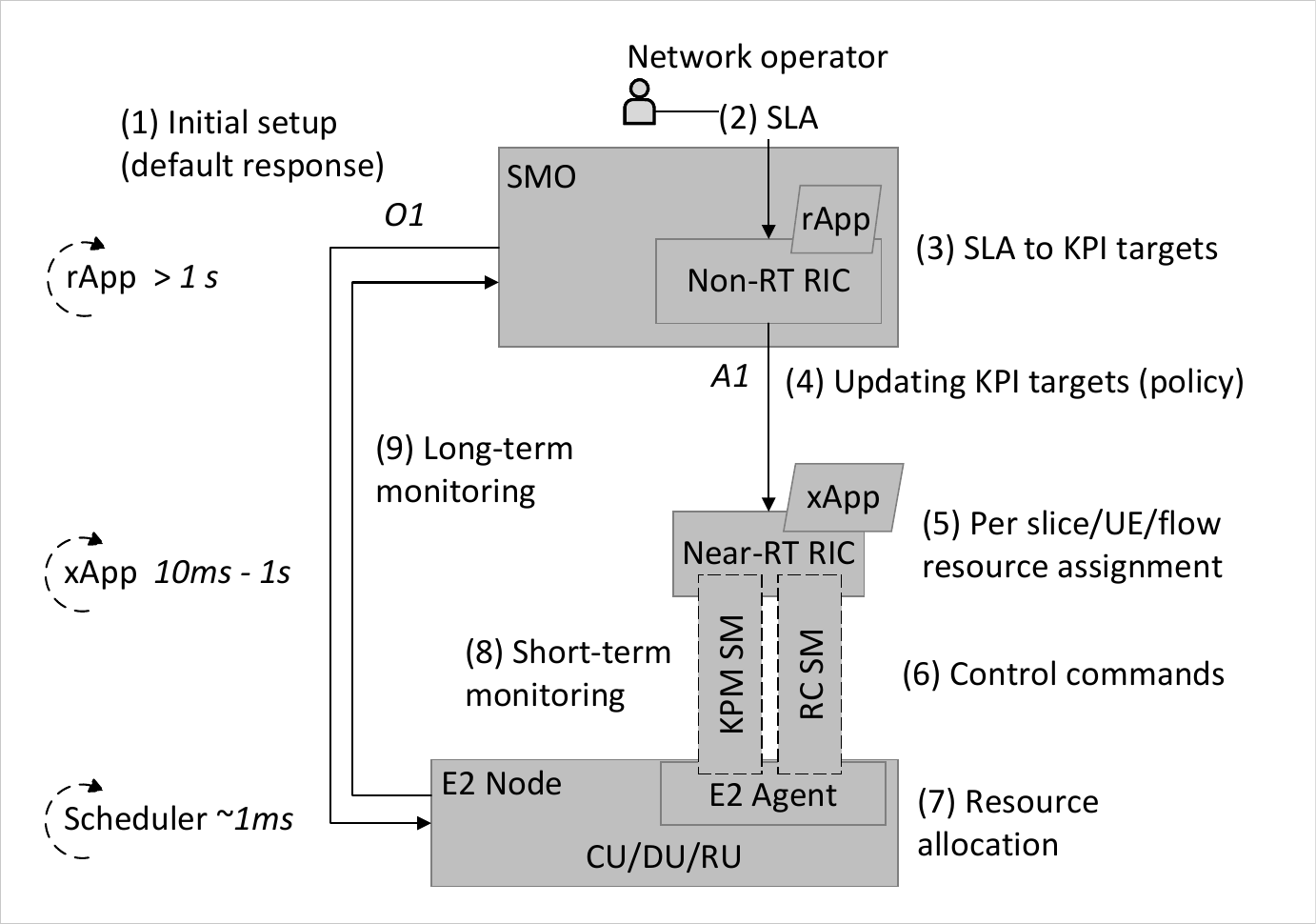}
\caption{{QoS operation in O-RAN}. The numbered steps exemplify the operation of QoS with network slicing and SLA assurance proposed by the O-RAN Alliance \cite{ORANWG12023UCASpecification, ORANWG22024WG2Usecases}. }
\label{fig:QoS-ORAN}
\end{figure}

 The operation of RAN slicing for QoS and SLA assurance proposed by the O-RAN Alliance is depicted in Fig. \ref{fig:QoS-ORAN} \cite{ORANWG12023Architecture, ORANWG32023RICARCH, ORANWG12023UCAReport, ORANWG12023UCASpecification, ORANWG22024WG2Usecases}. The SMO uses the O1 interface to provide {the} initial and default configuration of the slices (1), i.e., the proportion of resources allocated for a newly created slice, and the MAC scheduler that distributes the resources for each UE within the new slice. {Next}, the Non-RT RIC captures the service level agreements (SLAs) from the SMO (2). The Non-RT RIC is a subcomponent of the {Service Management and Orchestration (SMO)}, thus it has access to the data that is captured from the RAN ({through the} O1 interface; see (9) in Fig. \ref{fig:QoS-ORAN}). Using SLAs and the long-term performance metrics of the RAN, the Non-RT RIC is able to translate the SLAs to target KPIs (3). The latency and bit error rate are examples of target KPIs estimated by the Non-RT RIC to fulfill {the SLA} requirements of URLLC services. The Non-RT RIC can use {an rApp for the prediction} of events (such as traffic congestion) and optimized generation of target KPIs for different slices. The target KPIs are sent to the Near-RT RIC as A1 policies (4). The Near-RT RIC enforces these KPIs in the RAN (5). To achieve this, the Near-RT RIC uses an ML-based xApp that optimizes the assignment of resources to different slices according to the target KPIs and the RAN resource status (8). Finally, RAN control is enabled {for the RAN slice operation through the} E2 service models (7).

Currently, there are some efforts toward achieving the QoS framework presented in Fig. \ref{fig:QoS-ORAN}. Table \ref{tab1} shows existing studies that use O-RAN for QoS
in 6G RAN. There are three approaches: (1) Policy-level
management; (2) Resource allocation per RAN slice; and (3) Per-slice scheduling. 

 \begin{table}[!t]
\centering
\caption{{ Existing works for QoS in O-RAN developed by industry and research bodies.}}\label{tab1}
\begin{tabular}{p{1.5cm} p{1.0cm} p{1.3cm} p{1.3cm} p{0.5cm} p{0.5cm}} \hline
\multirow{2}{*}{\textbf{Work}} 
&\multicolumn{3}{c}{\textbf{Approach}}  
& \multirow{2}{*}{\textbf{DRL}}
& \multirow{2}{*}{\textbf{SLA}}
\\ \cline{2-4}
& \begin{tabular}[c]{@{}c@{}} (1) \\ Policy \\ level \end{tabular}
& \begin{tabular}[c]{@{}c@{}} (2) \\  Resources  \\ {per-slice} \end{tabular}
& \begin{tabular}[c]{@{}c@{}} (3) \\ 
{Per-slice} \\
scheduling \end{tabular}
&
&
\\ \hline
 2021 \cite{Bonati2021Intelligence} 
& -
& -
&  \large \cmark
& \large \cmark
& -
\\ \hline
2022 \cite{ZAMBIANCO2022Reinforcement}
& -
& -
& \large \cmark
& \large \cmark
& -
\\ \hline
2022 \cite{Papa2022Effects}
& -
& -
& \large \cmark
& -
& -
\\ \hline
 2022 \cite{Jian20225G}
& -
& -
& \large \cmark
& -
& -
\\ \hline
 2022 \cite{Lotfi2022Training}
& 
& \large \cmark
& \large \cmark
& \large \cmark
& - 
\\ \hline
 2022 \cite{Yang2022PlugFest}
& \large \cmark
& \large \cmark
& \large \cmark
& -
& \large \cmark
\\ \hline
 2022 \cite{KDDI2022PlugFest}
& \large \cmark
& \large \cmark
& \large \cmark
& -
& \large \cmark
\\ \hline
{2022} \cite{Zhang2022Joint} 
& \large \cmark
& -
& -
& -
& -
\\ \hline
 2023 \cite{RimedolabsOnfMwc2023QoS}
& -
& \large \cmark
& -
& -
& -
\\ \hline
 2023 \cite{Polese2023ColORAN}
& -
& \large \cmark
& \large \cmark
& \large \cmark
& -
\\ \hline
 2023 \cite{Tsampazi2023Comparative}
& -
& \large \cmark
& \large \cmark
& \large \cmark
& -
\\ \hline
 2023 \cite{Vladic2023PlugFest}
& \large \cmark
& -
& -
& -
& -
\\ \hline
 2024 \cite{Raftopoulos2024DRLBased}
& \large \cmark
& 
& \large \cmark
& \large \cmark
& \large \cmark
\\ \hline
 2024 \cite{Huang2024Validation}
& \large \cmark
& -
& -
& -
& -
\\ \hline
 2025 \cite{Dai2025ORANEnabled}
& -
& \large \cmark
& -
& \large \cmark
& - 
\\ \hline
 {2025} \cite{barker2025real}
& -
& \large \cmark
& -
& \large \cmark
&  -
\\ \hline
RSLAQ
& \large \cmark
& \large \cmark
& \large \cmark
& \large \cmark
& \large \cmark
\\ \hline
\end{tabular}
\end{table}

 \subsubsection{Policy-level management}
 Early approaches considered how to improve some of the mechanisms from the 3GPP QoS framework by using application domain information, user inputs, and edge services. The SDAP protocol that maps QFIs to DRBs has been first targeted in the design of applications that provide RAN control of UE-specific DRB-level QoS.  In this context, an xApp automates and optimizes the mapping of the QFI flows into the DRBs. For example,  \cite{Huang2024Validation} and \cite{Vladic2023PlugFest} present xApps that change the priority of the QoS flows of specific user equipment (UEs) according to the policies provided by the operator. In an emergency scenario, if two applications are running, e.g., video monitoring and file transfer streams, the operator elevates the priority of specific cameras by sending policies to the RIC by using the A1 interface (see (4) in Fig. \ref{fig:QoS-ORAN}).
 To support these approaches, the estimation or prediction of the QoS parameters, such as network-level (congestion), cell-level (throughput, delay, packet error, packet loss), and UE-level (throughput and latency performance), is of paramount {importance} \cite{Huang2024Validation} 
\cite{Kougioumtzidis2024QoE}. In this context,  the SMO collects long-term statistics from the RAN through the O1 interface (see (9) in Fig. \ref{fig:QoS-ORAN}), while the Non-RT RIC may employ ML-based applications (rApps) to make these predictions.

Optimization mechanisms to enhance traffic mapping between various flows have also been investigated using O-RAN features.
{The work in} \cite{Zhang2022Joint} {proposed} an optimization mechanism (mixed integer programming) that maps QoS flows into multiple routes by considering the retransmission of the packets. A single CU is connected to a pool of DUs. The DU pool contains a set of processors for packet retransmission scheduling. Under this approach, the delay time for URLLC traffic and {the} resource utilization (processors) for eMBB traffic are minimized. Similarly, the authors in \cite{Yassin2019Demo} experimented with a traffic prioritization mechanism to manage live video streaming (MBB) and remote car control (LLC). The mapping algorithm considers the state of each queue and the weighting of the allocation of radio resources according to the prioritized traffic volume.

Note that the approaches discussed in this section operate within the application and data planes, utilizing protocols of the 3GPP NR QoS framework, such as DRB formation. The following approaches represent efforts aimed at leveraging ML to optimize radio resource scheduling in the RAN according to the requirements of different applications and the network state.

\subsubsection{Resources per-slice}\label{sec:datadrivenQoS}
In this approach, an xApp is designed to optimize the distribution of the physical resource blocks (PRBs) per slice. For example, the work in \cite{Polese2023ColORAN} proposed the design of a DRL-based agent that allocates PRBs for three defined slices: eMBB, URLLC, and MTC. This approach maximizes the traffic rate for eMBB and the number of PRBs for MTC, and minimizes the buffer size for URLLC. Later, this work was extended by the authors in \cite{Tsampazi2023Comparative}. They reconfigured the reward function of the DRL by providing weights for each slice in order to avoid collisions between competing target KPIs.
{The work in}  \cite{barker2025real}  {experimented with the same approach in an open-source O-RAN testbed, emphasizing weighted traffic prioritization for URLLC, eMBB, and MTC.} 

\subsubsection{Per-slice {scheduling}}
In this approach, the xApp controls how the PRBs are distributed within each slice. A common technique is to find the optimal sequence of the MAC-scheduling algorithms (e.g., \textit{RR} and \textit{PF}) to apply in each slice based on network conditions. For example, in \cite{Polese2023ColORAN} and \cite{Bonati2021Intelligence}, the authors established 3 slices (eMBB, URLLC, and MTC) and three DRL agents, one per slice, that dynamically select the best scheduler at a given time.  As reported in these works, this dynamic selection significantly outperforms the individual operations of the schedulers. 

Another approach has been to replace the conventional MAC schedulers with optimized mechanisms.  For example, the work in \cite{ZAMBIANCO2022Reinforcement} proposed a DRL method that multiplexes the spectrum resources (PRBs) to optimize the cumulative throughput of eMBB and URLLC users. The mitigation of inter-numerology interference (INI) is considered in this method. The work in \cite{Papa2022Effects} studies scheduling policies for PRB allocation, including common mechanisms, such as MaxCQI, and other customized strategies. Their analysis also {introduced} centralized and distributed controllers and handover operations {to analyze} the effects on QoS. Although the distributed approach presents higher overhead and complexity in the control plane, the QoS is maintained compared to the single-controller approach.
In a similar study in \cite{Jian20225G} evaluated QoS-unaware and -aware (e.g., exponential/proportional fairness) MAC scheduling policies within three specific slices: VoIP (LLC), Video (eMBB), and CRB (mTC). As expected, delay-aware mechanisms {outperform unaware} methods in all scenarios. Note {that} \cite{ZAMBIANCO2022Reinforcement, Papa2022Effects, Jian20225G} assume direct assignment of PRBs from the xApp, which may be impractical due to the response time of xApps ($\geq$ 10\,ms).

All these previous works represent partial efforts to achieve the QoS operation presented in Fig. \ref{fig:QoS-ORAN}. Recent efforts have {attempted} to merge different strategies to achieve complete QoS management. For example, the authors in \cite{Polese2023ColORAN, Tsampazi2023Comparative} reported that running a single DRL agent to control the scheduling policy for slices and multiple DRL agents (one per slice) for spectrum resource scheduling in each slice significantly increases the performance with respect to the {5G NR base line schedulers}. On the other hand, Rimedo Labs and the Open Networking Foundation (ONF) created the QoS-based Resource Allocator (QRA) xApp, which follows A1 rules \cite{RimedolabsOnfMwc2023QoS}. These policies include equal distribution, reservation, and preference. The xApp allocates resources to various slices based on these rules and RAN parameters (e.g., SNR and throughput). However, the application of ML for resource distribution within each slice was not explored.

\subsubsection{SLA-driven QoS in O-RAN}
In a recent {demonstration,} \cite{Yang2022PlugFest} highlighted the importance of ensuring SLAs for UEs in specific slices. The policies are received from the operator through the A1 interface.  However, their work focused on a single type of slice aimed at guaranteeing the minimum throughput per UE. Our study extends this analysis to systems where multiple types of slices coexist, introducing challenges such as conflicts between SLA requirements of slices and priority management. Furthermore, \cite{Yang2022PlugFest} {did not} specify the use of ML. By incorporating intelligent methods such as DRL, we can better model system behavior and optimize resources, even under {varying conditions}.  A similar demonstration was conducted in \cite{KDDI2022PlugFest}, focusing on a single type of slice but analyzing different latency requirements. To guarantee the target delay for UEs, they reserved PRBs per slice and adjusted the UE/slice priority levels and packet delay budget. However, the use of ML was not specified in this work.

{Furthermore}, the authors in \cite{Raftopoulos2024DRLBased} showed that dynamic SLA requirements can be addressed using DRL. Although this work offered valuable insights for our approach, it focused on just one type of slice, which is a more simplified scenario compared to our method. Additionally, their study only considers per-slice SLAs and {does not address per-UE KPIs, which limits the granularity needed for 5G/6G applications.} The study in \cite{Lotfi2022Training} introduced an evolutionary DRL approach to address QoS for three slices: eMBB, MTC, and URLLC. While the approach was demonstrated in terms of maximizing QoS metrics—such as average data rate for eMBB, capacity for MTC, and delay for URLLC—it did not analyze SLA compliance. Furthermore, the approach involves directly assigning PRBs (overriding MAC scheduling) from the xApp, which might be impractical due to the response time of the xApp ($\geq$ 10\,ms), {as previously noted}. Finally, the authors in \cite{Dai2025ORANEnabled} presented a DRL-based solution for inter-slice resource allocation. The agent uses slice-level KPIs ({queue} length and head of buffer) to maximize data transmission and minimize delay.  They tested the solution using system-level simulations and an emulated testbed.  Unlike our solution, their DRL does not reflect SLA compliance or the network operator policy. 

Table \ref{tab1} summarizes the focus and contributions of the previous work. Our contribution is to optimize the allocation of resources both across slices and within individual slices while considering the SLAs specified by the Non-RT RIC in the form of target KPIs. To this end, we propose a DRL approach that incorporates the KPIs into the reward function.

\section{System architecture}\label{sec:3}
This section presents the proposed QoS framework for reliable SLA compliance, encompassing the workflow from operator intent to resource allocation across and within network slices.

\subsection{Intent-based SLA management}

\begin{figure*}[!t]
\centering
\includegraphics[width=\linewidth]{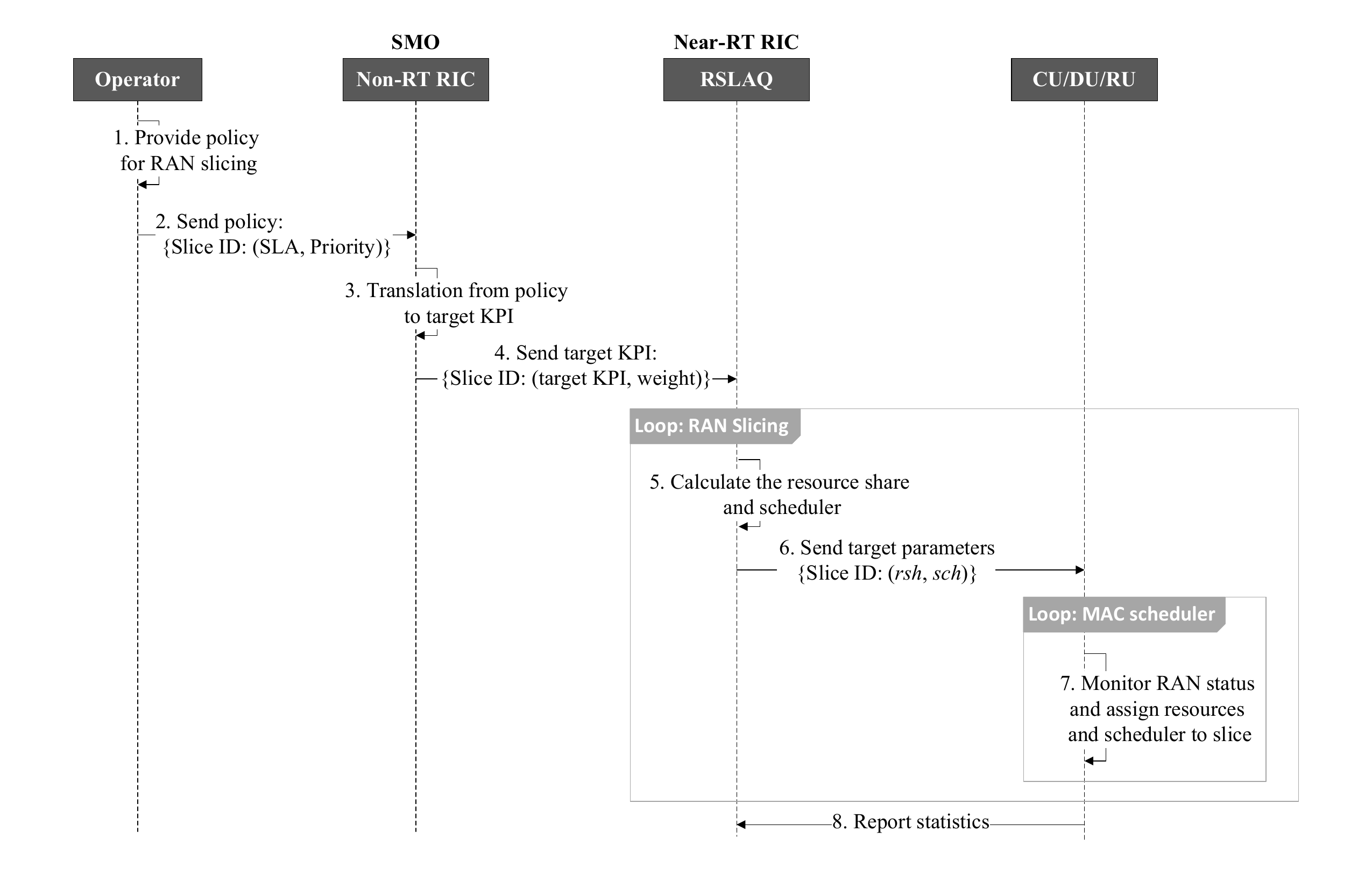}
\caption{{Intent-based SLA management}. The high-level policies of the network operator are translated and enforced into the RAN using the proposed framework.}
\label{fig:IntentSLAManagement}
\end{figure*}

Fig. \ref{fig:IntentSLAManagement} illustrates the operational workflow of the intent-based SLA management system. The network operator provides the parameters for the definition of the slices. Three types of slices are considered in O-RAN: {\textit{Reserved}, \textit{Policy}, and \textit{No-Policy}} slices~\cite{RimedolabsOnfMwc2023QoS, ORANWG22024A1TD}. A {\textit{Reserved}} slice receives a static quota of resources. \textit{{Policy}} and {\textit{No-Policy}} slices share the remaining resources, where the former adhere to SLAs and the latter are not bound by any policies. Although the design {of RSLAQ} provides flexibility to include the three types of slices, this study covers only {\textit{Policy}} and {\textit{No-Policy}} slices. DRL is used to optimally distribute resources between these slices.

 In Fig. \ref{fig:IntentSLAManagement}, the slice specifications for {\textit{Policy}} and {\textit{No-Policy}} slices are sent to the Non-RT RIC. In terms of {\textit{Policy}} slices, a set of slices is configured with different SLAs and priorities:  the lower the {value}, the higher the priority. For example, for URLLC, latency requirements are usually considered, whereas for eMBB, the achieved throughput {is taken into account}. A critical parameter in the slice configuration is the maximum number of UEs per slice, as this constraint enables consistent adherence to specified SLA requirements. In the URLLC example, if the number of users connected to this slice is very high, the system may not guarantee the latency requirements given its resource constraints. In the context of 6G applications, reliability in guaranteeing SLAs is also important and is defined {as the percentage of times the SLA requirements have been met}. Note that any conflicts between SLAs must be {resolved} in the Non-RT RIC before enforcing them in the system. The intent-level conflicts lie beyond the scope of this investigation.

The high-level specifications of SLA and priorities defined for the slices are translated by the Non-RT RIC to weights and target KPIs (see step 3 in Fig. \ref{fig:IntentSLAManagement}). Weights are calculated from the priorities and SLAs are translated into target KPIs that define the outage probability limits. These calculations are explained in detail in Section \ref{sec:4}.

In the following step (step 4 in Fig. \ref{fig:IntentSLAManagement}), the Non-RT RIC sends target KPIs and weights {to  RSLAQ}, which is {located} in the Near-RT RIC,  in the form of an A1 policy. Fig. \ref{fig:A1PolicyJson} shows an example of this A1 policy where the weights of the slices {are assigned} a proportional value to their provided priorities. Note that the sum of the weights must be 1. In addition, $\textit{outage kpis}$ and $\textit{soft kpis}$ (see definitions in Section \ref{sec:4}) are the main elements that are used by RSLAQ to estimate the resource distribution to be sent to the RAN components.

\begin{figure}[!t]
\input{json_A1_policy}
\caption{{Example of A1 policy.}}
\label{fig:A1PolicyJson}
\end{figure}

RSLAQ is designed to use DRL to calculate the resource share ($\textit{rsh}$) and RAN scheduler ($\textit{sch}$). To achieve this, it continuously monitors the statistics of the RAN and sends control commands to the RAN every  10\,ms (5G framework duration). The intents provided by the network operator are translated into numeric inputs at this layer. The two main control parameters sent to the RAN component are the distribution of resources (\textit{rsh}) and the scheduler to be used (\textit{sch}).

The MAC scheduler is responsible for enforcing the rules generated by RSLAQ (step 7 in Fig. \ref{fig:IntentSLAManagement}). Notably, the scheduler operates within its own control loop, which functions at 1\,ms intervals. The scheduler has the capability to override the control inputs of RSLAQ based on RAN metrics. For instance, if RSLAQ assigns a significant amount of transmission resources to an eMBB UE but the scheduler identifies that the user's transmission buffer is empty, the scheduler disregards the RSLAQ command. Instead, it redistributes the resources among other users while maintaining the proportionality dictated by RSLAQ. 

\subsection{RSLAQ design}

The core element of the proposed framework is the xApp that works as follows:  First, RSLAQ receives the target KPIs (SLA) from the Non-RT RIC through the A1 interface. Then, the xApp monitors the E2 Node (DU) to gather statistics. For the case study analyzed in this document, statistics of throughput (\textit{thr}), transmitted bytes (\textit{btx}), buffer status (\textit{bfs}), and resource share (\textit{rsh}) per UE are monitored. {In this study,} a single cell is considered. {RSLAQ} uses a DRL agent to select actions of resource proportion distribution ($p_j$) and scheduling selection (\textit{sch}). Finally, RSLAQ reports whether the target KPIs are being met and provides information on any failures, such as lack of resources. {The design of  RSLAQ} incorporates features that ensure robustness against variable user traffic patterns, manage resource conflicts during congestion events, minimize service outages, optimize resource utilization, and maintain slice isolation. A detailed architectural analysis is presented in Section \ref{sec:4}.

Note that the slice-aware scheduler is required on the E2 node. Furthermore, KPM and RC SMs are needed to enable monitoring and control of the E2 node. In the proposed design, the DRL agent captures metrics and executes actions every 10\,ms.

\subsection{Slice-Aware scheduler}
The proposed framework is designed for Time Division Duplex (TDD).  TDD operates with two independent schedulers for downlink (DL) and uplink (UL). The focus in this study is on the {DL}, which can be extended to UL. Furthermore, while 5G offers high flexibility in resource scheduling in frequency and time domains, our design aims to reduce computation and energy consumption by following {standard practice} and focusing on resource optimization at the frame level (defined as 10\,ms in this context).

The slice-aware scheduler works as follows: First, the TDD pattern is configured by the 3GPP 5G NR radio resource control (RRC) protocol. Subsequently, in the TDD pattern, slots (1\,ms each) designated for UL or DL are independently managed by the xApp commands.  In the DL slots, for example, the proportion of DRBs is distributed among the slices according to the xApp command. Finally, based on the selected scheduler (e.g., $RR$), resources within each slice per slot are allocated to users. This distribution is probabilistic rather than deterministic. For example, if three slices are considered and the distribution probabilities are [0.6, 0.2, 0.2], each time a PRB is considered by the scheduler, there is a 0.6 probability it will go to the first slice and a 0.2 probability it will go to the second or third slice. This probabilistic distribution is applied at the full frame level (10\,ms).

In terms of the MAC layer, the Hybrid Automatic Repeat Request (HARQ) protocol is activated to ensure reliable transmission over the air interface through error correction. Furthermore, the MAC scheduler prioritizes UEs with retransmissions (HARQ). However, only one retransmission is allowed per UE per Transmission Time Interval (TTI). Additionally, the scheduler allocates resources to UEs only if (i) there are remaining resources after handling retransmissions; (ii) the UEs have not received resources for retransmissions; (iii) the UEs have free HARQ processes; and (iv) the UEs have data ready to transmit. These conditions make the task of the DRL agent challenging, as the resulting states of actions are not deterministic. Nevertheless, reinforcement learning can effectively address non-deterministic scenarios, {which is demonstrated in our design.}

\section{Design of DRL agent}\label{sec:4}

\begin{table*}[h]
\centering
\caption{{SLA model for slices based on target KPIs ($m_1$ = eMBB, $m_2$ = URLLC, and $m_3$ = MTC). }}\label{tab:modelSLA}
\begin{tabular}{ p{1.5cm} p{1.5cm} p{6.3cm} p{7cm}} \hline
\textbf{Condition} 
& \textbf{Type}
& \textbf{Description} 
& \textbf{Example} 
\\ \hline
Optimization & Independent/ competing
& KPIs to optimize two or more slices that may be directly or indirectly competing. 
& $h_{1}^{1}$ = maximize throughput per UE for $m_1$, $h_{2}^{1}$ = minimize buffer occupancy per UE for $m_2$,  $h_{3}^{1}$ = maximize {throughput} per UE for $m_3$, $c_a$ minimize the cost of action. 
\\ \hline
\multirow{2}{*}{SLA} & 
Outage
& The target KPI is critical to fulfilling the required SLA.
& $k_{1}^1$ = minimum throughput for $m_1$, $k_{2}^{1}$ = maximum buffer occupancy per UE for $m_2$.
\\ \cline{2-4}
& Soft
& Target KPIs can be missed without consequences for SLA fulfillment, but they impact resource optimization.
& $k_{1}^{2}$ = maximum throughput per slice for $m_1$.
\\ \hline
\end{tabular}
\end{table*}

This section introduces the DRL agent designed to optimize resources and ensure SLA compliance, thereby, delivering QoS for demanding 6G applications.

\subsection{DRL formulation}

The design of the DRL agent is intended to be as simple as possible due to the response time requirements. Three important definitions are needed in this design: state, actions, and reward function. These definitions are presented below.

\subsubsection{State}\label{sec:4.4.1}
{Let $M$ be a set of slices such that $M=\{m_1, m_2, m_j, \dots, m_J\}$}. Also, $\mathcal{M}_j$ is the set of UEs assigned to slice $m_j$. The state of the network is defined based on the statistics gathered from the E2 node as follows:

\begin{eqnarray} \label{eq:state}
s & = & \begin{bmatrix} 
\textit{btx}_{m1} &  \dots & \textit{btx}_{mJ} & \textit{btx}_{cell} \\
\textit{bfs}_{m1} & \dots & \textit{bfs}_{mJ}  & \textit{bfs}_{cell}\\
\textit{rsh}_{m1} & \dots & \textit{rsh}_{mJ}  & \textit{rsh}_{cell}\\
\textit{tdp}_{m1} & \dots & \textit{tdp}_{mJ}  & \textit{tdp}_{cell}\\
\end{bmatrix},
\end{eqnarray}

\noindent where $\textit{btx}$, $\textit{bfs}$, $\textit{rsh}$ are the bytes to transmit, the buffer status, the share of resources, and (transmitted) dropped bytes, respectively. These parameters were selected to reflect the dynamics of the most common types of slices: \textit{btx} and \textit{bfs} help track eMBB and MTC performance, while \textit{bfs} and \textit{tdp} assist in monitoring {URLLC} performance. $s \in R^{4\times (J+1)}$, where $J$ is the number of slices. 

\subsubsection{Actions}\label{sec:4.4.2}
The DRL agent can control the proportion of PRBs per slice $p_j$ and the type of scheduler used inside the slices ($\textit{sch}$).  Note that in general $p_j = \{z: z \in \{0,1\} \text{ and } z \in R\}$.  Using this definition would result in a DRL agent with an infinite number of actions. To prioritize the speed of DRL learning and avoid unreasonable selections of $p_j$ (e.g., $\sum p_j>1$), we discretize and constrain $p_j$ as follows: Consider $J$ slices in the system. Let $p_j$ denote the proportion of PRBs for the $j$th slice. Therefore,

\begin{equation}\label{eq:p_jn}
  p_j = x | x \in  \{0, 0.1, 0.2, 0.3, 0.4, 0.5, 0.6, 0.7, 0.8, 0.9, 1\}.
\end{equation}

To introduce slice isolation in RSLAQ, we take a semi-dynamic approach to the distribution of radio resources in the RAN. We balance between the static and dynamic resource allocation strategies. In the static approach, a fixed number of resources are allocated to each slice, ensuring strong isolation but often leading to resource underutilization. Conversely, the dynamic approach dynamically distributes all resources without {a pre-defined slicing strategy}, enabling high resource optimization and flexibility but risking resource starvation due to the lack of isolation measures. RSLAQ operates as a hybrid solution, allocating 50\% of the resources statically while distributing the remaining 50\% dynamically in an optimized manner. The static component $p_\textit{sta}$ is distributed according to {weights assigned} by the network operator. The remaining 50\%, represented by $p_\textit{opt}$, is dynamically optimized using DRL. These components can be formally expressed as:

\begin{eqnarray}\label{eq:Psat_j}  
    p_\textit{sta}j & = & \omega_j\times50\% \textit{ of total PRBs} \\
     p_\textit{opt}j & = & \Theta(\cdot)\times50\% \textit{ of total PRBs},
\end{eqnarray}

\noindent where $\Theta(\cdot)$ represents the DRL model that optimizes the resources based on RAN statistics and SLA requirements. The proportion of PRBs applied for each slice is  

\begin{equation}\label{eq:p_j} 
    p_\textit{j} = p_\textit{sta} + p_\textit{opt}.
\end{equation}  

This approach achieves a balance between resource optimization and slice isolation. Table \ref{tab:RSLAactions} shows an example of slice isolation for three slices: eMBB, URLLC, and MTC.  $p_\textit{sta}$ takes values of [0.3333    0.4000    0.2667]$\times$ 50\% =   [0.1667    0.2000    0.1333]. The minimum proportion of PRBs to be assigned to each slice is that given by $p_\textit{sta}$. Note that if the slice is a no-policy slice, $p_\textit{sta}$ is 0. The maximum proportion of resources to be provided to a slice is the total of PRBs in the system minus the proportions that have been statically assigned to the other slices. Finally, $p_\textit{opt}$ can take a value between 0 to 1, as it is multiplied by 50\% corresponding to the amount of resources assigned to the optimization part.
 
\begin{table}
\centering
\caption{Example of definition of actions for slice isolation.}.
\begin{tabular}{|p{1.0cm}| p{1.3cm} | p{1.8cm} |p{1.1cm}|p{1.0cm}|}
\hline 
\textbf{Slice} & $\omega$ (priority) & Min $p_j$  = $p_{\textit{sta}}$  & max $p_j$&  $p_{\textit{opt}}$ \\
\hline 
eMBB & 0.3333 (2) & 0.1667 & 0.6667 & (0,1)\\
\hline
URLLC &  0.4000  (1) & 0.2000 &0.7& (0,1) \\ \hline 
MTC &  0.2667 (3) & 0.1333& 0.6333 & (0,1)\\
\hline
\end{tabular}
\label{tab:RSLAactions}
\end{table}

In terms of the scheduler, consider the scheduling options for the slices as

\begin{equation}
  \textit{sch} = \{\textit{RR}, \textit{PF}, \textit{BCQI}\}.
\end{equation}

These three schedulers have been selected because they are among the most well-known and widely used in resource distribution. This approach maintains compatibility with current base station operations, requiring no modifications to their native scheduling capabilities.

Finally, the set of actions $A$ is as follows:

\begin{equation}\label{eq:A}
 A  =  \{(\text{perm}(\text{perm}\{p_1, \dots ,p_j, \dots, p_J\}:\sum_{j=1}^J p_j=1), \textit{sch})\}.
\end{equation}

\subsubsection{Reward function for radio resource optimization}
The resources are optimized using the equation below.
\begin{equation}\label{eq:rop}
    r_{\textit{opt}} = \sum_j \omega_j h_j + \dfrac{1}{\text{cost}(\alpha)},
\end{equation}

\noindent  where $h_j$  represents the KPI to optimize for each slice $j$. $\omega_j$ allows to prioritize a slice if two or more slices have competing KPI targets. Examples of optimization targets are provided in Table \ref{tab:modelSLA}. $\text{cost}(\alpha)$ represents the cost associated with the action. In this study, the cost of the scheduler is

\begin{equation}\label{eq:cost} 
    \text{cost}(sch) = \begin{cases} 1 & \mbox{if } \textit{RR}  \\ 
    2 &  \text{otherwise}, \end{cases} 
\end{equation}  

\noindent where the agent receives higher reward with $\textit{RR}$ because it does not require CQI reports, making it less costly in terms of frequency, computation, and energy compared to $\textit{PF}$. $\textit{RR}$ can be more effective in scenarios where users have similar channel conditions.

We incorporate SLA requirements into DRL learning to enable intent-based operations, ensuring optimal resource utilization for the operator and QoS guarantees for users. In our approach, we model SLAs as target KPIs and integrate them into the reward function.

For each slice $m_j \in  M$, there is a set of target KPIs $K_{j}=
\{k_{j}^1, k_{j}^2, \dots, k_{j}^i, \dots \}$, where $k_{j}^i$ represents the $i$th target KPI for slice $j$. The SLA violation rate ($\textit{vrsla}_j$) is defined as the number of UEs that experience SLA violations divided by the total number of users that belong to that slice:

\begin{equation}\label{eq:vrsla}
     \textit{vrsla}_j^i  = \dfrac{P(\text{KPI}<k_{j}^i)}{ \lvert\mathcal{M}_1\lvert}
\end{equation}

Note that the requirements for some target KPIs may be more critical than for others. 
In this context, $K_{out}$ is the set of target SLAs that are critical for fulfillment of the SLAs.  On the other hand, while KPIs in $K_{soft}$ do not impact SLA fulfillment, they are important for optimizing resource utilization. For instance, failing to meet the maximum throughput limit per slice for eMBB (see $k_1^2$ in Table \ref{tab:modelSLA}) can lead to unnecessary additional resource allocation, resulting in QoS that exceeds the agreed SLA.
While this is advantageous from the user perspective, it is less favorable from the operator standpoint, as the remaining resources could be better allocated to other services.

The outage condition for a slice can be defined in terms of the reliability parameter provided by the network operator. Using \eqref{eq:vrsla}, an outage condition for slice $m_j$ is defined as

\begin{equation}\label{eq:psij}
    \varphi_j=\begin{cases}
1,  & \mbox{if } 
   \exists \textit{ vrsla}_j^i \text{ such that }  \textit{vrsla}_j^i > 1-\textit{reliability}_j \\ 
0,  & \mbox{otherwise,} 
\end{cases}
\end{equation}

\noindent where $\varphi_j \in \{0,1\}$ is a binary value representing an outage condition in slice $m_j$. Similarly, the condition of soft SLA violations is defined by $\rho_j$, using $k_{j}^i \in K_{soft}$. The reward function for SLA assurance is defined as follows:

\begin{equation}\label{eq:reward}
r_k=\begin{cases}
0 \text{ (Terminal $s$),} & \mbox{if }  \sum_j \rho_j > 0 \\
-\sum_j(\varphi_j w_j) \text{ (Terminal $s$),}  & \mbox{if } 
    \sum_j \varphi_j > 0 \\ 
r_{opt},  & \mbox{otherwise,} 
\end{cases}
\end{equation}
 
\noindent where the first and second elements represent the outage probability and soft probability, respectively, indicating the failure to achieve any outage or soft target KPI. $r_{opt}$ in Eqn. \eqref{eq:reward} is defined in Eqn. \eqref{eq:rop} and optimizes the resources per slice and per UE within each slice.

\subsection{Formulation for 3 slices: eMBB, URLLC, and MTC}
The definitions of actions, states, and reward are provided for the case study of 3 slices: eMBB, URLLC, and MTC.  The state in Eqn. \eqref{eq:state} is given as $s \in R^{4\times 4}$. The action set is defined in Eqn. \eqref{eq:A}, where $A  = \{a_1, a_2, \dots \}$. For three slices, an example of this action is $a_1 =  (p_1,  p_2,  p_3, RR)$ where $p_j$ is given by \eqref{eq:p_j}. The number of possible actions is  $|a| = 198$, and thus $A \in R^{198 X 4}$.   

From the examples presented in Table \ref{tab:modelSLA}, the KPIs for eMBB, URLLC, and MTC slices are grouped as follows:  target KPIs for eMBB are $\{k_{1}^{1}, k_{1}^{2}\}$ and the optimization KPI is $\{h_1^1\}$; {the} target KPI for URLLC is $\{k_{2}^{1}\}$ and the optimization  KPI is $\{h_2^1\}$; and the optimization KPI for MTC is $\{h_{3}^{1}\}$.

The  reward component of outage KPIs in Eqn. \eqref{eq:reward} is

\begin{multline}\label{eq:rou}
  r_{out}=
    P(\text{thr per slice for } m_1 < k_{1}^{1})
 \text{ or } \\ P(\text{bfs per UE for } m_2 > k_{2}^{1}).
 \end{multline}

The  reward component of soft KPIs in Eqn. \eqref{eq:reward} is

\begin{equation}\label{eq:rsoft}
r_{soft}=
    P(\text{maximum thr per slice for } m_1 > k_{1}^{2}).
\end{equation}

Finally, $r_{opt}$ in Eqn. \eqref{eq:reward} for this 3-slice example is given as

\begin{eqnarray}
   \textit{{h}}_{1}^1  & = & \omega_{1} \dfrac{1}{\lvert \mathcal{M}_1 \lvert}\sum_{\text{UE } \in   \mathcal{M}_1}{thr(\text{UE})}, \label{eq:rc1}\\ 
     \textit{{h}}_{2}^1   & = &{\omega_{2}}  e^{-\left(\max_{\text{UE } \in  \mathcal{M}_2}{bfs(\text{UE})} \right)}, \label{eq:rc2} \\
     \textit{{h}}_{3}^1   & = & \omega_{3}    \dfrac{1}{ \lvert \mathcal{M}_3 \lvert}\sum_{\text{UE } \in   \mathcal{M}_3}{thr(\text{UE})}, \label{eq:rc3}\\
   c_{\alpha} & = & \text{cost}(sch)\label{eq:rc4}.
\end{eqnarray}

All components in Eqn. \eqref{eq:rc1} - \eqref{eq:rc4}  are positive. Thus, if a KPI requires minimization, such as for uRLLC, it is converted to a positive value through an exponential function. Additionally, all elements must be normalized (e.g., we use maximum achievable rates for the throughput normalization). Weights $\omega_j$  allow the definition of priorities per slice according to the configurations of the network operator. In our experiments, we consider priorities of [2 1 3] for eMBB, URLLC, and MTC set by the operator, which translates to $\omega_j = [0.33, 0.40, 0.27]$.

\subsection{DRL algorithm}
Dual Q-learning (DDQL) with experience replay is used to solve the optimization problem. This solution is chosen for its simplicity and timely response. Algorithm \ref{algorithm:deepQLearningExperienceReplay} presents the DDQL implementation for RSLAQ. DDQL uses two deep neural networks (DNNs) with the same structure: An Online DNN with parameters ${\theta}$ (weights and biases)  and a Target DNN with parameters $\hat{\theta}$. While the Online DNN ($\theta$) is trained in each learning step to decrease the loss function,  the Target DNN ($\hat{\theta}$) is frozen to enhance learning stability. The Target DNN ($\hat{\theta}$) is periodically updated to match the Online DNN (${\theta}$) after a pre-defined number of learning steps. The experience replay mechanism stores past experiences in a dataset with entries of the form $\Braket{s, \alpha, r, s'}$ (current state, action, reward, next state). Furthermore, mini-batches of this dataset are used to train the DNNs, which helps the agent reduce the number of interactions that are required to learn and reduce the variance in learning updates.

\begin{algorithm}
\begin{algorithmic}[1]
\Require $\epsilon$, $\lambda_\epsilon$, $\epsilon_{\textit{min}}$, $\gamma$, $E$, $L$, $\textit{btsz}$, $\textit{nsut}$,  $\textit{ntsr}$
\Ensure Best action {$\Braket{p_j,\textit{sch}}$} for each state 
 \State Initiate Online DNN ($\theta$) and Target DNN ($\hat{\theta}$)
 \State Initiate replay memory $V$ to capacity $L$
 \State Set initial state $s$
 \For{\textit{steps=1} to $E$}
  \State      Choose 
       \begin{equation*} 
        \alpha = \begin{cases} \mbox{random} & \mbox{if } \epsilon  \\ \mbox{max}_\alpha Q(s,\alpha:\theta) &  \mbox{if } {1-\epsilon} \end{cases} 
        \end{equation*}  
    \State     SET\_ACTION($\alpha$)   \Comment{Execute the action}
    \State $[\textit{tstate}, r]$ = GET\_REWARD($\textit{stats}$, $K_{\textit{out}}$, $K_{\textit{soft}}$, $\textit{reliability}$, $\alpha$)  
    \State Observe $s'$
    \State    Store transition $\Braket{s, \alpha, r, s', \textit{tstate}}$ into $V$
        \If {${\textit{steps}}>\textit{btsz}$ } 
        \State Retrain\_QNetwork($V$, $L$)
        \State $\epsilon$ = SET\_EPSILON\_DECAY($\epsilon$)  \Comment{Simulated annealing}
        \EndIf
        \If{$\textit{mod}(\textit{steps} , \textit{nsut}) = 0$}
         \State Update $\hat{\theta}$ of the Target NN with $\theta$ of the Online NN
        \EndIf
        \If{$\textit{mod}(\textit{steps},\textit{ntsr})==0$}
         \State $\textit{tstate} = 1$ \Comment{Reset to initial state}
        \EndIf
        \State Set $s=s'$ \Comment{Move to the new state}
 \EndFor
\State  Use ﬁnal ${\theta}$ to retrieve the action $\alpha$ with the highest Q-value for each state $s$
 \end{algorithmic}
 \caption{DDQL with experience replay}  \label{algorithm:deepQLearningExperienceReplay}
\end{algorithm}

Algorithm \ref{algorithm:deepQLearningExperienceReplay} receives the exploration parameter $\epsilon$, the exploration decay rate $\lambda_\epsilon$, the minimum exploration decay rate $\epsilon_{\textit{min}}$, the discount factor $\gamma$, the number of learning steps $E$, the size of the replay memory $L$, the minibatch size $\textit{btsz}$ ($\textit{btsz} <= L$), the number of steps before updating the Target DNN $nsut$, and the number of steps before the environment is reset $\textit{ntsr}$. The output is the best action ($p_j$ and $\textit{sch}$) for each state. 

The algorithm first initializes the Online and Target DNNs with  ${\theta}$ and $\hat{\theta}$, respectively, and the replay memory $V$ to capacity $L$ (Lines 1-2). The agent goes through a ﬁnite number of $\textit{steps}$ in a loop until it reaches a maximum of steps $E$ (Lines 4-21). In our design, a learning episode comprises a sequence of $\textit{steps}$ that correspond to the states between an \textit{initial state} and a \textit{terminal state}. A \textit{terminal state} is defined either through the reward function (outages; see Lines 27 and 30) or after every $\textit{ntsr}$ steps. Each $\textit{step}$ consists of selecting and performing an action, changing the state, and receiving a reward. 

In Lines 5-20, the agent selects an action $a \in \mathcal{A}$ by using the $\epsilon$-greedy method (Line 5). Thereafter, the agent executes the selected action $a$ in the environment (Line 6), computes the reward $r$, and gets the next state $s'$ (Line 8). Then the agent stores the tuple $\Braket{s, a, r, s', \textit{tstate}}$ in the replay memory $V$ (Line 9). The variable \textit{tstate} is a binary indicator for the terminal state, where 1 indicates a terminal state and 0 indicates a non-terminal state.

Once the replay memory contains more $\textit{btsz}$ experiences (Line 10), the agent randomly takes a mini-batch (with size $\textit{btsz}$) from the replay memory $V$ to train the Online DNN ($\theta$) (Line 11). In this process, for all samples in the mini-batch, the agent predicts
$\textit{target}$ by using the Online DNN ($\theta$) and the Target DNN ($\hat{\theta}$). After that, the agent uses gradient descent and backpropagation algorithms to adjust the weights and biases  of the Online DNN, $\theta$, and to minimize the $\textit{Loss} = (\textit{target} - Q(s,\alpha))^2$.

After the execution of $\textit{nsut}$ steps (Lines 14-15), the agent updates the weights and biases of the Target NN, $\hat{\theta}$, with the weights and biases of the Online NN, ${\theta}$. Finally, the agent uses the ultimate $\theta$ to retrieve the best actions with the highest Q-values for each state (line 22).  

We heuristically explored different hyperparameters and chose the ones that delivered the best performance as follows:  {$E = 300$,} $L=500$, $\gamma = 0.85$, $\epsilon=0.1$, $\textit{nts}=300$, $\textit{nsut}=20$. The Online and Target DNNs have the following structure:  four 2D convolutional layers, each followed by a batch normalization layer, and a fully connected layer at the output. The layers are using the Tanh activation function.

\section{Experimental results}\label{sec:5}

For the numerical simulations, we used MATLAB version 24.1 along with 5G Toolbox and Communications Toolbox. Additionally, we employed the Wireless Network Simulation Library\footnote{\url{https://uk.mathworks.com/matlabcentral/fileexchange/119923-communications-toolbox-wireless-network-simulation-library}}. Currently, MATLAB lacks support for 5G NR layers above the RLC, i.e., RRC, PDCP, and SDAP. Therefore, functionalities such as handover are not available. The simulations were carried out on a high-performance computing cluster, utilizing 4 nodes with 64 tasks per node and a memory allocation of 8 GB per CPU.

We first setup a 5G scenario with one gNB and 3 slices (eMBB, URLLC, and MTC). This configuration covers the specification of the target scenarios in this study, where multiple services co-exist with varying requirements. We evaluate the performance of the DRL agent with new, unseen traffic and compare the results to those of traditional schedulers, focusing on SLA fulfillment.

\subsection{Experimental setup}
A highly detailed system-level simulation was employed to evaluate the solution.  Table \ref{tab:experimentParameters} shows the most relevant experimental parameters used in the numerical simulations. We consider a bandwidth of 50 MHz and numerology 1 ($\mu$=1,  subcarrier spacing of 15\,kHz, 50 PRBs). These PRBs are shared between three slices: {eMBB}, URLLC, and MTC. The {TTI is} 1\,ms (1 slot), i.e., there are 10 slots per 5G frame. The TDD periodicity is 5\,ms, and the pattern is described in Table \ref{tab:experimentParameters}. Note that the xApp decision is per 5G frame resolution, but the data can be captured at a higher sampling frequency. In terms of application layer, the slices serve a total of 20 UEs: 5 UEs for eMBB, 5 UEs for URLLC, and 10 UEs for MTC. The operator configures which user belongs to which slice. 

\begin{table}
\centering
\caption{Parameters used in the numerical simulations}
\begin{tabular}{p{1.3cm} p{2.2cm} p{2.3cm} p{1.2cm}}
\hline
 \textbf{Component}
& \textbf{Parameter}
& \textbf{Value}
& \textbf{Unit}
\\  \hline
\multirow{6}{*}{gNB}
& Carrier Frequency  
& 2.59e9 (n38)
& Hz
\\  \cline{2-4}

& Duplex Mode
& TDD
& -
\\ \cline{2-4}

& Channel bandwidth
& 10e6
& Hz
\\
\cline{2-4}

& Subcarrier spacing 
& 15e3
& Hz
\\ \cline{2-4}
& NumRBs
& 50
& RBs
\\ \hline
\multirow{2}{*}{TDD}
& Frame 
& 10
& ms
\\ \cline{2-4}
& DLUL periodicity
& 5
& ms
\\ \cline{2-4}
& DLUL pattern
& D$\rvert$D$\rvert$8D$\rvert$4GB$\rvert$4U$\rvert$U$\rvert$U
& Slot
\\ \hline
\multirow{4}{*}{Scheduler}
& CellNumber
& 1
& -
\\ \cline{2-4}
& RBAllocationLimit
& 50
& RBs/TTI/UE
\\ \cline{2-4}
& Scheduling type
& Slot-based
& -
\\ \cline{2-4}
& Scheduler
& Slice-aware
& -
\\ \hline
\multirow{3}{*}{\begin{tabular}[l]{@{}c@{}} App  \\ layer \end{tabular}}
& Slices 
& eMBB, URLLC, MTC
& -
\\ \cline{2-4}
& Number of UEs
& 20 (5 for eMBB, 5 for URLLC, and 10 for MTC)
& -
\\ \cline{2-4}
Physical Layer
& Full PHY with perfect channel estimation.
& (SISO) antenna
& -
\\ \hline
\end{tabular}
\label{tab:experimentParameters}
\end{table}

Table \ref{tab:exp_sla_networkconditions} shows five experimental scenarios tested with RSLAQ, where each scenario challenges the robustness of RSLAQ as follows:

\begin{enumerate}
    \item {\textit{Low traffic}}: In this scenario, the downlink traffic for eMBB users is low, while the minimum \textit{thr} per slice configured for this slice is 10\,Mbps. RSLAQ must guarantee the assignment of resources to comply with this target KPI, but at the same time, if the eMBB resources are not used, they must be distributed between MTC and URLLC users.
    \item {\textit{Normal}}: {This scenario is expected during standard operation of the network}; therefore, RSLAQ must be able to respond according to the SLAs provided by the operator, i.e., maintaining the target KPIs and priorities.
    \item {\textit{Congestion}}: {This scenario is expected to occur regularly with 6G applications}. In this experiment, the traffic of eMBB and MTC users is very high. RSLAQ must be able to prioritize and optimize the use of the resources to fullfill the SLAs by solving conflicts between them.
    \item {\textit{Stressed}}: This is the most challenging scenario that RSLAQ can face, where, in addition to network congestion, the SLA provided by the operator has stringent requirements. 
    {To represent this, the minimum
     \textit{thr} for eMBB users is elevated compared to previous experiments.}
    \item {\textit{Insufficient resources}}: The design of RSLAQ assumes that the SLA must align with the resource allocation established in a previous stage at the Non-RT RIC (SMO). However, it is important to analyze the {behavior of RSLAQ} in this scenario. This allows us to generate safeguard mechanisms based on RSLAQ outputs, considering that there could be instances where the Non-RT RIC can misconfigure the SLA requirements with respect to the available resources. {These mechanisms contribute to the overall system robustness.}
\end{enumerate}

\begin{table*}[ht]
\centering
\caption{Experimental conditions: varying network conditions and SLAs.}
\begin{tabular}{ p{1.3cm} p{4.3cm} p{11.0cm} }
\hline
Experiment
& Network condition
& SLAs
\\ \hline
{\textit{Low traffic}}
& eMBB= 50\,Kbps;
URLLC= 1\,Mbps;
MTC= 2\,Mbps;
& eMBB=\{min \textit{thr} = 10\,Mbps; max \textit{thr} = 15\,Mbps;  Priority = 2, Maximize \textit{thr}\};
URLLC=\{max \textit{bfs} = 3\%; Priority = 1, Minimize \textit{bfs}\};
MTC  =\{No Policy ; Priority = 3; Maximize \textit{thr}\}
\\ \hline
{\textit{Normal}}
& eMBB= 70\,Mbps;
URLLC= 1\,Mbps;
MTC= 2\,Mbps;
& eMBB=\{min \textit{thr} =10\,Mbps; max \textit{thr} = 15\,Mbps;  Priority = 2, Maximize \textit{thr}\};
URLLC=\{max \textit{bfs} = 3\%; Priority = 1, Minimize \textit{bfs}\};
MTC=\{No Policy; Priority = 3; Maximize \textit{thr}\}
\\ \hline
{\textit{Congestion}}
& eMBB= 100\,Mbps;
URLLC= 1\,Mbps;
MTC= 100\,Mbps
& eMBB=\{min \textit{thr} = 10\,Mbps; max \textit{thr} = 15\,Mbps;  Priority = 2; Maximize \textit{thr}\};
URLLC=\{max \textit{bfs} = 3\%; Priority = 1; Minimize \textit{bfs}\};
MTC=\{No Policy; Priority = 3; Maximize \textit{thr}\}
\\ \hline
{\textit{Stressed}}
& eMBB= 100\,Mbps;
URLLC= 1\,Mbps;
MTC= 100\,Mbps
& eMBB=\{min \textit{thr} = 20\,Mbps; max \textit{thr} = 25\,Mbps;  Priority = 2, Maximize \textit{thr}\};
URLLC=\{max \textit{bfs} = 3\%; Priority = 1; Minimize \textit{bfs}\};
MTC=\{No Policy; Priority = 3; Maximize \textit{thr}\}
\\ \hline
{\textit{Insufficient resources}}
& eMBB= 100\,Mbps;
URLLC= 2\,Mbps;
MTC= 100\,Mbps
& eMBB=\{min \textit{thr} = 20\,Mbps; max \textit{thr} = 25\,Mbps;  Priority = 2; Maximize \textit{thr}\};
URLLC=\{max \textit{bfs} = 3\%; Priority = 1; Minimize \textit{bfs}\};
MTC=\{No Policy; Priority = 3; Maximize \textit{thr}\}
\\\hline
\end{tabular}
\label{tab:exp_sla_networkconditions}
\end{table*}

\subsection{DRL training}

\begin{figure}[!t]
    \centering    \includegraphics[width=\linewidth]{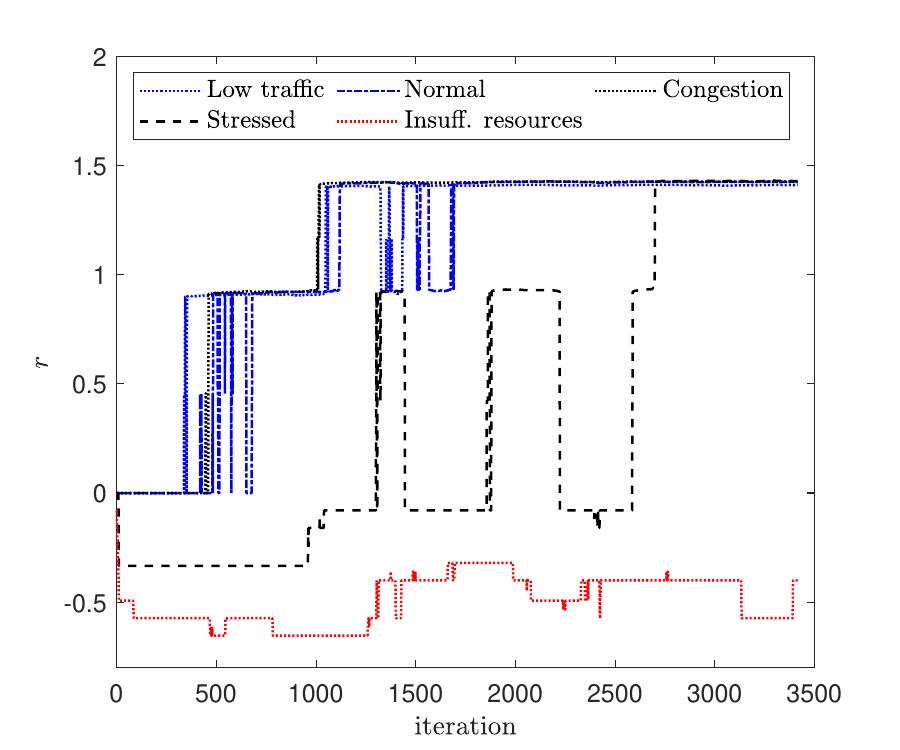}
    \caption{{Reward function of RSLAQ for different SLAs and network conditions.}}
    \label{fig:reward_RSLAQ}
\end{figure}

Fig. \ref{fig:reward_RSLAQ} shows the average reward in the learning iterations for all scenarios. In general, in scenarios 1-4, after an initial exploration phase, the system successfully converges to a solution that minimizes outages and {soft KPI violations}. Scenarios 1-4 are incrementally complex to solve in terms of network conditions and SLAs, and RSLAQ shows robustness to learn in these different scenarios. {In scenario 4 (\textit{stressed})}, which combines network congestion and stringent SLA requirements,  RSLAQ finds it difficult to enforce the SLA policy provided by the operator. Nevertheless, it achieves optimal conditions after a number of iterations.

The reward function is designed to produce negative values if RSLAQ cannot {fulfill outage SLAs}.
Therefore, if the radio resources are insufficient to cover SLA specifications, RSLAQ reward values are negative (see {\textit{Insufficient resources}} in Fig. \ref{fig:reward_RSLAQ}). In particular, we observed that in our {experiments for} scenario 5,  the traffic for URLLC is very high and the available resources {cannot meet} the minimum \textit{bfs} per URLLC UE, even when {RSLAQ} allocates all available resources (50 PRBs) exclusively to this slice.  If this event is detected, RSLAQ sends an alarm to the Non-RT RIC, which should promptly correct the misconfiguration. In the mean time, RSLAQ can revert to a previous configuration or {any safety} condition established by the operator to ensure the QoS for users is  maintained.

\subsection{DRL testing}

\begin{figure*}[!t]
\centering
    {%
        \includegraphics[width=\textwidth]{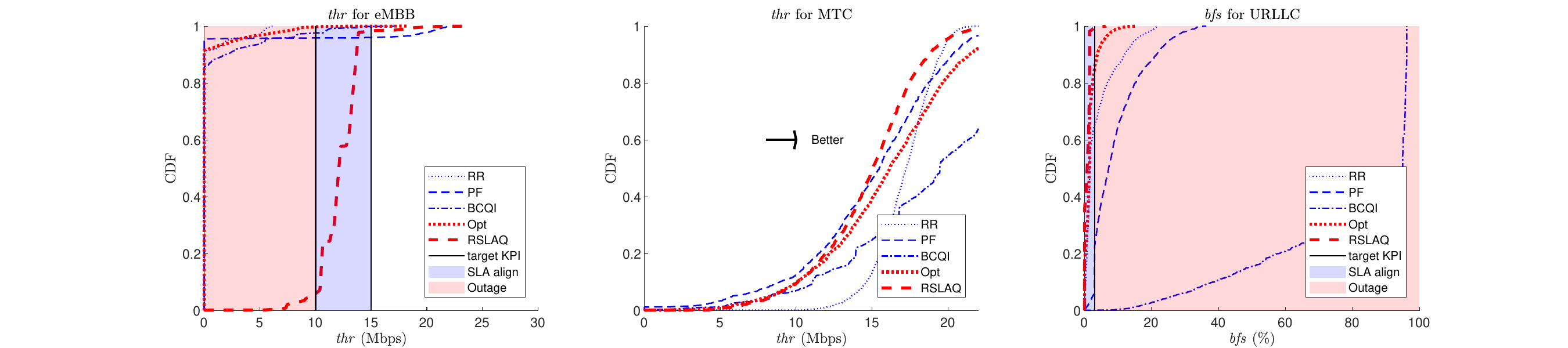}
        \label{fig:rewarda}
    } 
\parbox{0.4\textwidth}{ \centering (a) Low traffic} 
    \hfill
    {%
        \includegraphics[width=\textwidth]{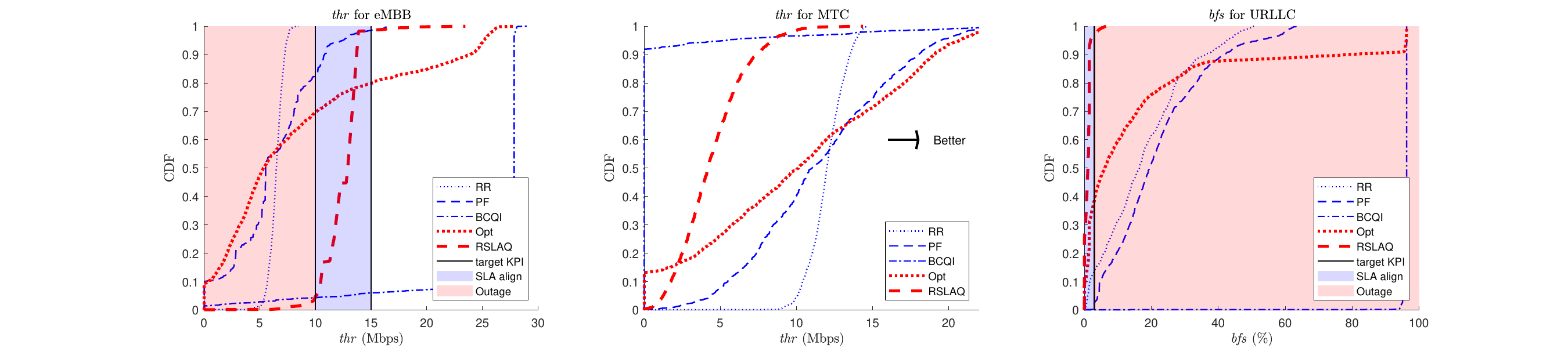}
        \label{fig:rewardb}
    } 
 \parbox{0.4\textwidth}{ \centering (b) Normal traffic}
    \hfill
    
    {%
        \includegraphics[width=\textwidth]{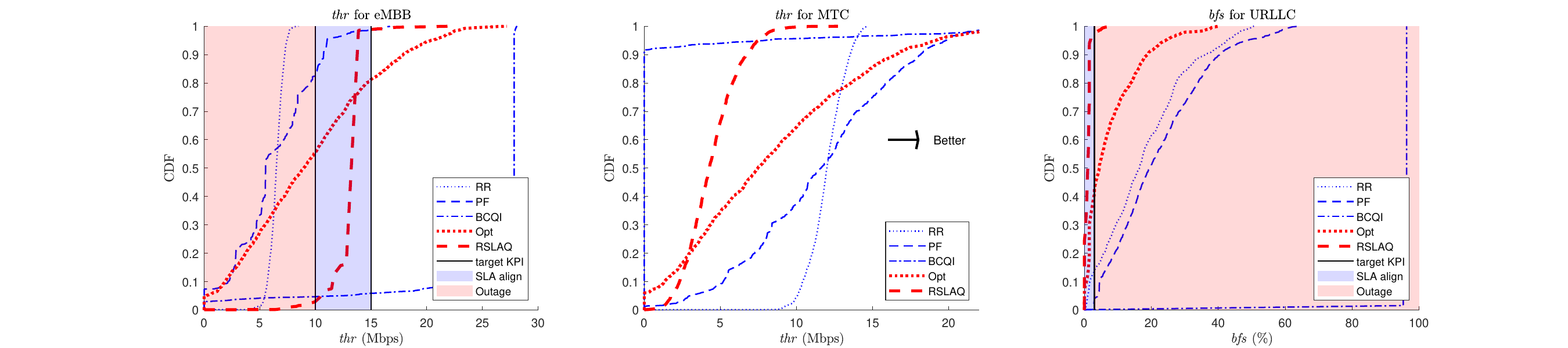}
        \label{fig:rewardc}
    } 
 \parbox{0.4\textwidth}{ \centering (c) Congestion}
    \hfill

    {%
        \includegraphics[width=\textwidth]{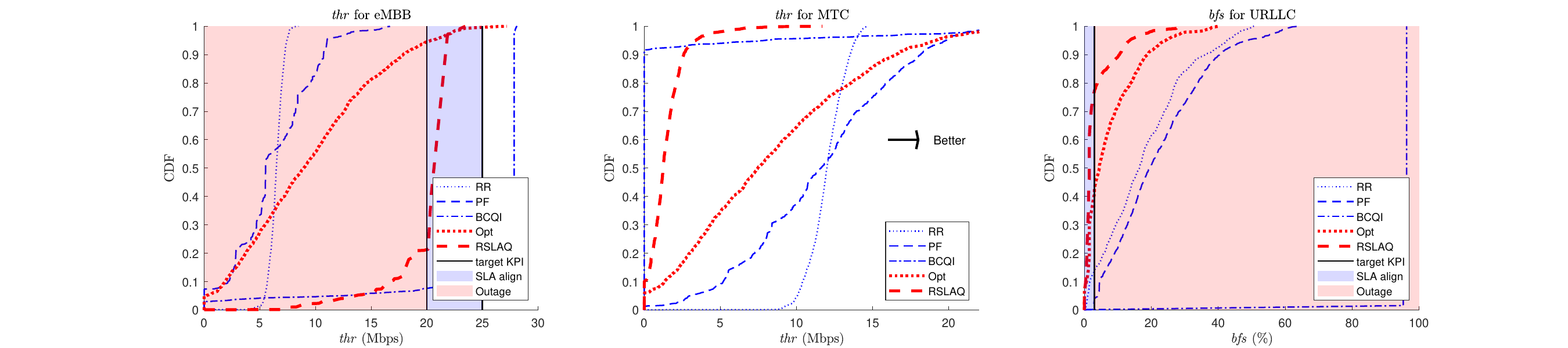}
        \label{fig:rewardd}
    } 
 \parbox{0.4\textwidth}{ \centering (d) {Stressed}}
    \hfill

    {
        \includegraphics[width=\textwidth]{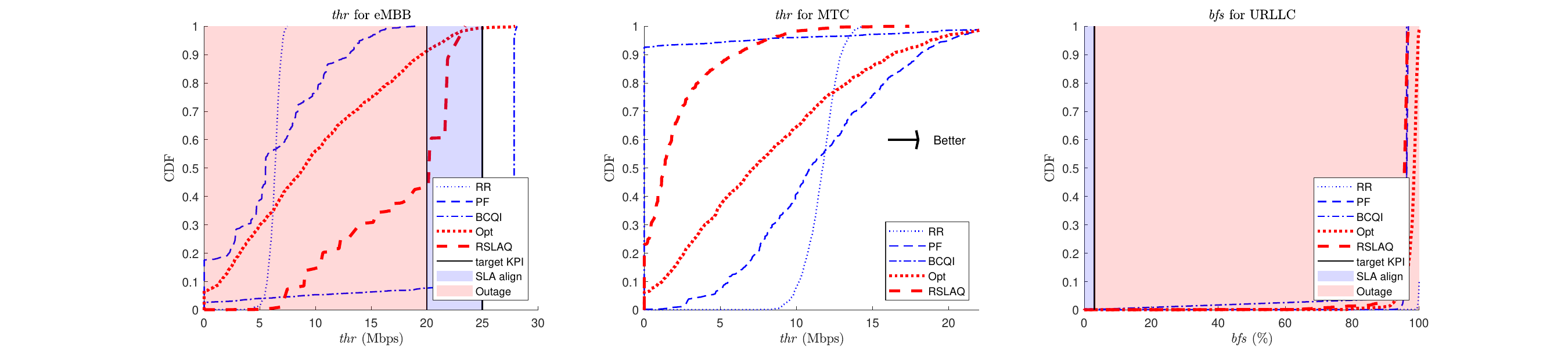}
        \label{fig:rewarde}
    } 
     \parbox{0.4\textwidth}{ \centering (e) Insufficient resources}
     \hfill
    \caption{RSLAQ performance vs. other mechanisms.  RSLAQ enforces target KPIs (SLAs) in different traffic and SLA conditions considered in the design (sufficient resources). When the system resources are insufficient, URLLC cannot be fullfiled.}
\label{fig:SLAs_comparison}
\end{figure*}

Fig. \ref{fig:SLAs_comparison} compares the performance of standard schedulers (\textit{RR}, \textit{PF}, and \textit{BCQI}), an optimization approach, and RSLAQ. The optimization approach (Opt) reflects previous work aimed at maximizing resources per slice, as in \cite{Tsampazi2023Comparative}. In the first scenario (Fig. \ref{fig:SLAs_comparison} (a)), RSLAQ allocates the resources following SLA specifications for eMBB and URLLC and achieves maximum \textit{thr} for MTC. Other approaches cannot comply {with target KPIs}, particularly with eMBB. Note that RSLAQ allocates resources for eMBB, but since the traffic for eMBB is low (see Table \ref{tab:exp_sla_networkconditions}, first row) these resources are allocated to URLLC and MTC users. In this scenario, the double control loop in our framework  helps {allocate} unused resources.

In the second scenario, in Fig. \ref{fig:SLAs_comparison} (b), the users of all slices behave as expected {under normal} conditions (high throughput for eMBB, a higher number of devices with low throughput for {MTC} and low throughput for URLLC). Unlike other mechanisms, RSLAQ is able to optimally distribute the resources to fulfill SLAs.  If SLAs are not considered, while some KPIs may show improvement with certain schedulers, they often come at the cost of other important KPIs. For example, in the case of the optimized approach shown in Fig. \ref{fig:SLAs_comparison} (b), the \textit{thr} for MTC is significantly better than that of RSLAQ. However, under these conditions, the outage KPI for URLLC is violated. 

The third scenario ({\textit{congestion}}) in Fig. \ref{fig:SLAs_comparison} (c) {is also frequent in 5G/6G}. In this scenario, the downlink traffic for MTC and eMBB users is very high. Prioritization comes into {play} in this scenario, where MTC has the lowest priority, and URLLC the highest priority. RSLAQ properly allocates the resources and selects the schedulers to fullfil the SLAs, as seen in Fig. \ref{fig:SLAs_comparison} (c).

In the {\textit{stressed}} scenario (see Fig. \ref{fig:SLAs_comparison} (d)), the robustness limits of RSLAQ {are} tested, as in addition {to network} congestion, the SLAs are very strict. In particular, eMBB target KPIs are increased. Under these conditions, there are instances where outages cannot be avoided, as the resources cannot cover {all SLAs} at the same time. RSLAQ gives priorities to which outages should be minimized, and in our experiments, it is the URLLC. {The weights} in $\omega_j$ provide flexibility {in controlling} the behavior of RSLAQ in this scenario. As seen in Fig. \ref{fig:SLAs_comparison} (d), {the outage} SLAs for eMBB and URLLC are minimized and the resources allocated to MTC are very low.

Finally, we also present the results for {the scenario in which the} resources are insufficient to fulfill outage SLAs in Fig. \ref{fig:SLAs_comparison} (e). In this scenario, the target KPI for URLLC cannot {be met} with the resource available in the system. RSLAQ cannot guarantee {an optimal resource} distribution in this scenario, as the allocation of resources to the system (50 PRBs) cannot be modified at the Near-RT RIC level. RSLAQ logs this event and notifies the Non-RT RIC which should reconfigure the system in terms of the resource allocation and SLAs.

Overall, as shown in Figs. \ref{fig:SLAs_comparison} (a)-(d), the optimized approach (Opt.) outperforms \textit{RR}, \textit{PF}, and \textit{BCQI} by simultaneously maximizing \textit{thr} for eMBB and MTC while minimizing \textit{bfs} for URLLC (considering  priorities $\omega_j$). However, this approach still fails to meet all SLAs. RSLAQ not only optimizes resource allocation, but also ensures that the distribution aligns with SLAs. Our solution demonstrates that connecting the application plane (user policy) and the data plane (channel) through the intelligent QoS xApp in the O-RAN RIC significantly improves resource utilization, benefiting both the users and the network operator.

\section{Discussion, limitations, and future work} \label{sec:6}

This section highlights the flexibility, robustness, and reliability benefits of RSLAQ, explores potential improvements, addresses its limitations, and suggests directions for future work.

\subsection{Robustness, reliability, and scalability}

In terms of robustness, we have tested four different scenarios (varying network conditions and SLAs) for which RSLAQ demonstrated consistent performance in SLA fulfillment. Note that the design of the reward function is {self-explanatory} and enables rapid assessment of whether the system has reached a feasible solution or is unable to find one. If the long-term reward remains below zero at the end of the training, no feasible solution was identified. Note that this scenario only occurs if the available resources and SLAs are not properly validated at the Non-RT RIC. RSLAQ incorporates {fail-safe} mechanisms that include the activation of default 3GPP schedulers, reverting the system to a previous condition, or using default slicing strategies (e.g., reserved slices) to maintain system performance until RSLAQ is {re-trained} with different configurations of resources and SLAs.  

\begin{figure*}[!t]
    \centering
\includegraphics[width=\linewidth]{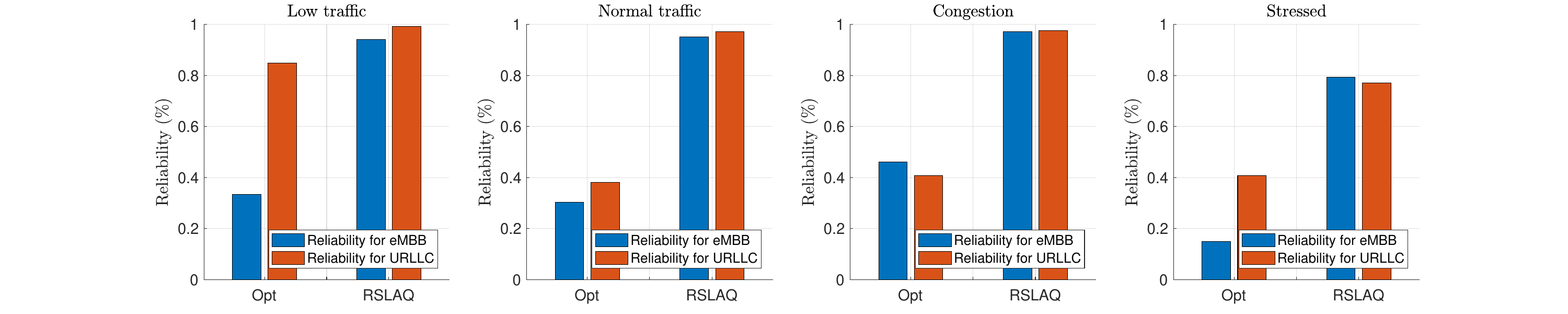}
\caption{{Outage probability of RSLAQ vs. the optimized approach.}}
\label{fig:reliability_comparison}
\end{figure*}

A key requirement for 6G communications is high reliability, especially critical for URLLC. Fig. \ref{fig:reliability_comparison} compares the average reliability of RSLAQ, calculated as $(1 -P(k_{\textit{out}}))$, with the optimized approach (Opt). The standard schedulers \textit{PF}, \textit{RR}, and \textit{BCQI} have been excluded from this analysis, as they are QoS-unaware. Furthermore, the scenario of insufficient resources is not considered, as RSLAQ operates under the assumption that the Non-RT RIC guarantees that the existing resources cover the SLA requirements.

As shown in Fig. \ref{fig:reliability_comparison}, RSLAQ achieves the highest simultaneous reliability for eMBB and URLLC. For the first three scenarios, which are expected to cover most of the network conditions, RSLAQ achieves a high reliability rate ($>$ 95\%). This value is reduced to approximately 80\% when the network {is congested} and the operator has allocated resources and SLAs at their maximum allowed limits. 

 In terms of scalability, RSLAQ can scale to any number of UEs. In addition, when the number of slices increases (i.e., $>3$), the design of RSLAQ can be flexibly adapted as explained next. The number of actions increases proportionally to the number of slices. Since the action space defines the number of neurons in the output layer of the DNN in the DRL architecture, a change in the number of slices changes the DNN design. In terms of the reward function, as new slices are defined, new SLAs appear. The calculation of the reward is based on the SLAs and thus the reward function should include them. With respect to performance, {the demands on memory, processing resources, and learning time increase as more scenarios need to be evaluated}. Nevertheless, the runtime performance of the DNN is expected to be maintained once the model has been trained. With these modifications, offline re-training is required before the model is uploaded to the system. A very high number of slices may hinder DRL learning. Coarse-grained discretization of $p_j$ can help reduce the complexity of the learning process. 
 
\subsection{Limitations and future work}

\subsubsection{Enhancements to RSLAQ}
To achieve a more robust and faster learning convergence of the DRL, the design necessarily limited the action space, particularly the resolution of the proportion of PRBs/slice. As shown in Fig. \ref{fig:reliability_comparison}, the DRL successfully met the performance objectives despite this low resolution.  Lower resolution reduces resource needs for model training and deployment, resulting in lower energy consumption. Nevertheless, depending on the computational capability available, this resolution can be increased to enhance performance. However, note that increasing the resolution significantly increases the complexity of the neural network. For example, using the 0.01 resolution results in $A \in \mathbb{R}^{15453 X 4}$, making the size of $A$ 78 times larger than with a 0.1 resolution. This design could result in longer learning times and possibly limited exploration of certain action-state spaces due to the infinite number of possible states. Note that considering a continuous action with $p_j \in \mathbb{R}$ may be unsuitable for systems that require high reliability.  In the RSLAQ design, we prioritized fast learning and high reliability.

\subsubsection{Additional SLAs}
Integrating an admission control mechanism (e.g., handovers) could enhance the ability of the system to manage SLAs and accommodate additional SLA options, such as the maximum number of UEs per slice. Furthermore, although we do not explicitly include an energy model in the system, the efficient distribution of PRBs contributes to better energy utilization. Currently, RSLAQ cannot control transmission power.  The implementation of these features requires protocols at the RRC layer and functionalities within the Non-RT RIC, which are beyond the scope of this study.  These aspects can be explored in future research. 

\subsubsection{Security of ML in O-RAN}
The use of ML in O-RAN introduces new vectors for adversarial attacks~\cite{Yungaicela2024Misconf}. The softwarization and openness of O-RAN facilitate the deployment of rogue elements. When deploying RSLAQ in real-world scenarios, robust authorization and authentication mechanisms for UEs, E2 nodes, and third-party xApps must be {employed} to prevent intrusions that could severely compromise the performance of the system. For instance, rogue UEs could introduce misleading channel state information (CSI), compromising the normal operation of the schedulers and negatively affecting the learning and performance of the DRL agent. Additionally, malicious xApps might corrupt KPMs stored in the RIC database (such as $\textit{thr}$, $\textit{bfs}$, and UEs/slice), undermining the inputs of the DRL method and thus compromising SLA compliance. Future research will focus on developing reactive (e.g., anomaly detection) and proactive (e.g., adversarial learning) defense mechanisms to ensure the robustness of DRL systems against such adversarial attacks.

\section{Conclusion}\label{sec:7}

A reliable DRL-based QoS xApp, RSLAQ, was designed and tested using detailed simulations.  Our design using 3GPP standard mechanisms, such as using $RR$, $PF$, and $BCQI$ schedulers, allows seamless integration of RSLAQ {into} existing 5G/6G systems without significant modifications. Experimental results show that incorporating QoS awareness into the RAN system not only maximizes KPIs for each type of service but also ensures compliance with specific SLAs. { RSLAQ presented consistent performance across a diverse range of scenarios, demonstrating its robustness and adaptability in varying network conditions.} This approach benefits both the {users} by maintaining their QoS and the network operator, as it enables the deployment of additional services without compromising the agreed {quality of existing ones.}

\section*{Acknowledgments}
This work is supported through the \textcolor{black}{NICYBER2025 programme funded by Innovate UK. The ORANSecAI project is a collaboration with Ampliphae.} The views expressed are those of the authors and do not necessarily represent the project or the funding agency.

\bibliographystyle{IEEEtran}
\bibliography{references}

\vfill

\end{document}